\documentclass[english,aps,pra, twocolumn, floatfix, showpacs,  notitlepage]{revtex4-1}
\usepackage{lipsum}
\usepackage{amsmath}
\usepackage{amsfonts}
\usepackage{graphicx}
\usepackage{color}
\usepackage{hyperref}
\usepackage{epstopdf}
\begin{document}
\title{Pattern formation in driven condensates}
%

%


%
 \author{Ivana Vasi\'c}
 \email[Corresponding author: ]{ivana.vasic@ipb.ac.rs}
\affiliation{Institute of Physics Belgrade, University of Belgrade, Pregrevica 118, 11080 Belgrade, Serbia
}
\author{Du\v san Vudragovi\'c}
\affiliation{Institute of Physics Belgrade, University of Belgrade, Pregrevica 118, 11080 Belgrade, Serbia
}
\author{Mihaela Carina Raportaru}
\affiliation{Institute of Space Science – INFLPR Subsidiary, 409 Atomistilor, Magurele, Romania}
\author{Alexandru Nicolin-\.{Z}aczek}
\affiliation{Institute of Space Science – INFLPR Subsidiary, 409 Atomistilor, Magurele, Romania}
\begin{abstract}

The onset of pattern formation in a spatially homogeneous system subjected to external driving is an important topic in various scientific fields. A celebrated classical example is the Faraday instability, where a vertically oscillated fluid surface undergoes a parametric resonance, giving rise to standing waves that self-organize into regular spatial patterns.
Bose–Einstein condensates (BECs) provide an ideal quantum-mechanical platform for studying pattern-forming mechanisms due to their exceptional degree of experimental control. As a compressible state of quantum matter, a condensate responds sensitively to external perturbations, including time-periodic modulation of interactions, trapping potentials, or external fields. These features make BECs particularly well suited for exploring driven nonequilibrium phases and pattern formation.
In this chapter, we review the remarkable progress achieved in this field over the past two decades. We begin with the first theoretical proposal predicting parametric instabilities and emergent Faraday waves in driven condensates. We then discuss key experimental and theoretical breakthroughs that confirmed these predictions and refined the understanding of the underlying mechanisms. This line of research has culminated in the recent observation of a stabilized square lattice pattern in a periodically driven BEC condensate confined in a two-dimensional geometry. This driven superfluid state with superposed density modulation was shown to exhibit some features of a supersolid state.

{\it
This chapter is dedicated to the memory of Antun Bala\v z, whose guidance as a mentor, contributions as a colleague, and friendship were deeply valued (amicus certus in re incerta).
}

\end{abstract}

\maketitle


\section{Introduction}
\label{sec:introduction}

Systems of cold atoms represent a versatile experimental platform with easily tunable relevant parameters of a quantum system \cite{RMP.80.885}. On top of this, many of these parameters can be modulated in time. One of the most studied states of cold atoms, both theoretically and experimentally, is a Bose-Einstein condensate (BEC) \cite{book_of_Pethick_Smith}. Properties of this state typically depend on the external trap parameters and types and strength of atomic interactions. Since BEC is a compressible state of matter, by modulating external parameters excitations are created, that can further be probed through precise measurements.  All these features and excellent experimental possibilities make BECs a perfect playground for studying pattern formation under external driving, which is a topic of great interest in various scientific disciplines \cite{RMP.65.851}.

Pattern formation is a phenomenon observed in a variety of systems, ranging from physical and chemical to biological and ecological domains, in which a uniform state spontaneously breaks into an ordered structure due to internal interactions or external driving forces. Prominent examples include reaction-diffusion systems, where chemical species undergo interactions leading to stationary or oscillating patterns, famously described by Turing \cite{Turing1952}.
 In biological systems, pattern formation is seen in processes like animal coat patterns \cite{Nature.376.32}. In physics, motivation for studying pattern formation under driving often comes from the classical phenomenon of Faraday waves, first studied and described by Michael Faraday at the beginning of  19th century~\cite{PhilTransRSocLond.121.299}.
It was discovered that when the container with the liquid is harmonically oscillated in a vertical direction, the wave patterns may emerge on the surface, depending on the ratio of the liquid depth and the container size, as well as depending on the modulation frequency.  The interest for this type of excitations arose again during the 1980s, as a consequence of the study of nonlinear liquids. Similar phenomena have been observed in diverse systems, ranging from classical fluids~\cite{Phys.Rev.Lett.52.922, Phys.Rev.Lett.78.4043, J.Fluid.Mech.805.591, Phys.Fluids.30,Exp.Fluids.61, Proc.R.Soc.Lond.452}, to multimode lasers~\cite{Phys.Rev.Lett.80.3968}, and superfluid helium~\cite{PhysRevE.76.046305}.

In the context of ultracold gases, Faraday waves were first investigated theoretically in 2002~\cite{PhysRevLett.89.210406} for a quasi-two-dimensional BEC. The first Faraday waves were measured in an elongated BEC experiment with $^{87}$Rb in 2007~\cite{PhysRevLett.98.095301}, and afterwards in experiments with $^{7}$Li~\cite{PhysRevA.81.053627,PhysRevX.9.011052} and $^{23}$Na~\cite{PhysRevLett.121.185301, PhysRevLett.128.210401}. Faraday waves were also experimentally investigated for a fermionic superfluid of $^6$Li \cite{NJP.23.103038}.
In the mentioned experiments  the radial part of the harmonic trap was modulated, except for the case of $^7$Li where the contact interaction strength was  modulated. However, qualitatively, this leads to the same type of dynamics.
Early theoretical studies focused on systems with local interactions \cite{PhysRevLett.89.210406,PhysRevA.76.063609,PhysicaA.389.4663,ProcRomAcad.12.209,RomRepPhys.63.1329,RomRepPhys.65.820}. Subsequently, Faraday waves were studied in dipolar~\cite{PhysRevA.81.033626,PhysRevA.86.023620,ProcRomAcad.14.35} and two-component condensates, including systems with spatially-dependent contact interaction~\cite{PhysRevA.85.023613,PhysRevA.89.023609, JPhysB.49.165303, PhysicaA.391.1062, Phys.Rev.A.100.063610, Phys.Rev.A.105.063319}. Numerical studies of Faraday waves were also extended to spin-1 BECs~\cite{arXiv:2510.08087}, to mixtures of Bose and Fermi gases~\cite{PhysRevA.87.023616},  as well as Fermi gases exhibiting superfluid behavior~\cite{PhysRevA.78.043613,JPhysB.44.115303}.
While these previous studies focused on harmonic traps and box-potentials \cite{NatPhys.17.1334}, more recently ring-shaped condensates have been shown to provide an alternative geometry for exploring non-linear pattern formation~\cite{Phys.Rev.A.99.023619, Phys.Rev.A.99.053619, Phys.Rev.X.8.021021}.
Pattern stabilization in BECs in two dimensions attracted a lot of attention  and several experiments investigated this process \cite{ NatPhys.16.652, Phys.Rev.Lett.127.113001, PhysRevA.109.L051301,PhysRevX.15.011026, arXiv:2505.17409}. Finally, it was discovered that some driven two-dimensional BECs may even possess supersolid-like phase characteristics as in the recent experiments of Refs.~\cite{NatPhys.21.1064, NatPhys.21.1036}.  Harmonic driving of the scattering length was also proposed as a mechanism to stabilize a BEC with attractive interactions \cite{EPJD.65.106} and to control excitations of a BEC \cite{PhysRevA.84.013626, PhysRevA.85.033635}. It was also found that deep optical lattices suppress the onset of Faraday waves \cite{PhysRevA.83.013603}.

Faraday waves in ultracold gases are a consequence of the existence of parametric resonances in the system. Often, BEC dynamics is well described by the time-dependent Gross-Pitaevskii equation \cite{book_of_Pethick_Smith} which is a starting point in many of the theoretical and numerical studies of Faraday waves. By assuming a weak driving amplitude and using some approximations, it is possible to derive an effective Mathieu equation \cite{book_of_McLachlan} that exhibits the phenomenon of parametric resonance. From this equation it is then possible to describe the growth rate of an unstable mode and spatial and time periodicity of emergent Faraday waves. This description works well for quasi-one-dimensional or elongated BECs. However, in two dimensions this approach captures only the short-time dynamics and to describe pattern stabilization it is necessary to consider amplitude equations.

In addition to Faraday waves, superfluids of cold atoms exhibit other types of instabilities. In recent experiments counter flow instabilities of two counterpropagating superfluids  were investigated \cite{NatPhys.20.939, NatPhys.21.1398}. Typical of this process is the deformation of an initial  vortex sheet and the onset of a flutter-finger pattern. The Rayleigh-Taylor instability was observed in a mixture of two immiscible BECs \cite{SciAdv.11.9752} exposed to a uniform force acting in opposite directions on the two species. This instability results in the onset of mushroom-like patterns at the initially flat interface between the two species.

In the more general context of cold atoms, periodic driving is of great importance. Through the mapping known as Floquet engineering \cite{PhysRevX.4.031027}, driving provided a way to enrich the set of quantum states realized with cold atoms \cite{RMP.89.011004}. In particular, artificial gauge fields were realized in this way. The downside of this approach is the eventual thermalization under driving and a transition to a featureless, infinite-temperature state. However, long-enough times  of  a few hundred milliseconds have been achieved in experiments where a relevant ``prethermal'' Floquet state properly describes the system \cite{RMP.89.011004}.

In the following we first review some basic properties of the above discussed parametric resonance. We then describe the experimental and theoretical consensus of Faraday waves in an elongated BEC with short-range interactions, which is the most thoroughly studied case of Faraday waves in BECs. Afterwards we summarize some of the theoretical results showing how long-range dipolar interactions enrich the phenomenon. We then indicate the progress toward studying Faraday patterns in higher-dimensional settings by describing the onset of surface waves in a pancake shaped BEC, stabillization of different lattices under bichromatic driving as well as recent results on the stabilization of a square-lattice pattern.

\section{Methods and general considerations}
\label{sec:methods}
 BEC dynamics is well described by the nonlinear Gross-Pitaevskii (GP) equation \cite{book_of_Pethick_Smith, RMP.71.463}
\begin{equation}
 i\hbar\frac{\partial\psi}{\partial t}=\left[-\frac{\hbar^{2}}{2m}\nabla^{2}
+V(x, y, z, t)+g(t) N\left|\psi\right|^{2}\right]\psi,
\label{eq:GPE}
\end{equation}
where $\psi(x, y, z, t)$ denotes the BEC wave function (normalized to unity),
$N$ represents the number of atoms and $g(t)=4 \pi \hbar^2 a_s(t)/m$ is
proportional to the
scattering length $a_s(t)$ of the interatomic interactions, and $m$ is the atomic mass.
The external
potential confining the atoms is typically described by a harmonic trap
\begin{eqnarray*}
V(x,y, z, t) & = &
\frac{1}{2}m\omega_{r}^{2}(t)\left(x^{2}+y^2\right)+\frac{1}{2}m\omega_{z}^{2}z^{2},
\end{eqnarray*}
where $\omega_r=\omega_x=\omega_y$ represents the radial trap frequency and $\omega_z$ is the longitudinal confinement frequency.
In cold atoms one way to induce patterns is by the harmonic modulation of the radial trap strength
\begin{equation}
 \omega_{r}(t) = \omega_r(0) \,(1+\varepsilon \cos(\omega\, t)),
 \label{eq:modulated_trap}
\end{equation}
where $\omega$ is the driving frequency  and $\varepsilon$ is the driving amplitude.
Another approach is to exploit the Feshbach resonance technique \cite{Nature.392.32354, RMP.82.1225} and modulate the interaction strength
\begin{equation}
  g(t) = g(0)\,(1+\varepsilon \cos(\omega\, t)).
  \label{eq:modulated_interaction}
\end{equation}
Note that often in the actual implementation, the driving term in the two last equations can be given by $\sin(\omega \,t)$ instead of $\cos(\omega\, t)$.
Direct numerical simulations of the full Eq.~(\ref{eq:GPE}) can successfully describe the onset of patterns under driving \cite{PhysRevLett.89.210406, PhysRevA.76.063609, PhysRevA.85.023613, PhysRevA.89.023609, PhysRevX.9.011052, PhysRevLett.121.243001, JPhysSocJpn.92.064602}.
In particular, properties of Faraday patterns in an elongated, cigar-shaped BEC
are properly described by a full numerical simulation of Eq.~(\ref{eq:GPE}) as well as by using a linear stability analysis of Eq.~(\ref{eq:GPE}) \cite{PhysRevA.76.063609, JPhysSocJpn.92.064602}.
A closely related approach was used for the understanding of surface patterns observed in two-dimensional BECs \cite{Phys.Rev.Lett.127.113001}.
More recently, stabilization of a square lattice in a two-dimensional BEC in a specially designed trap potential under driving given by Eq.~(\ref{eq:modulated_interaction}) was observed and explained starting with Eq.~(\ref{eq:GPE}) \cite{PhysRevA.109.L051301, NatPhys.21.1064, PhysRevX.15.011026}.

In numerical simulations, an initial state is usually selected as the ground state of the BEC. The ground state is typically obtained by imaginary-time propagation of Eq.~(\ref{eq:GPE}). A possible density modulation is seeded either by inherent numerical noise or by explicitly adding noise to the initial wave function \cite{Phys.Rev.Lett.127.113001}. A systematic numerical study \cite{JPhysSocJpn.92.064602} has shown that the choice of the dynamical protocol - e.g. modulating either the trap potential as in Eq.~(\ref{eq:modulated_trap}) or the contact interaction as in Eq.~(\ref{eq:modulated_interaction}) does not play a key role for the properties of the emergent patterns.

For weak driving amplitudes, linearization of the GP equation can give meaningful results. However, as the driving amplitudes get stronger nonlinear effects, such as mode coupling and nonlinear shifts in the frequencies of collective modes, become more pronounced \cite{PhysRevA.84.013618, PhysRevLett.98.095301}.
Finally, for very strong driving amplitudes, the initial BEC might get fragmented.
This occurred in the experiment of Ref.~\cite{PhysRevX.9.011052}, as strong driving amplitudes were used to probe the onset of patterns for low driving frequencies. Fragmented BEC is beyond the description given by Eq.~(\ref{eq:GPE}) and to describe the onset of the observed granulation the multiconfigurational time-dependent Hartree method was used in Ref.~\cite{PhysRevX.9.011052}. It was shown that the state obtained by driving exhibited pronounced quantum fluctuations and correlations.

\begin{figure}[tbhp!]
          \includegraphics[width=0.45 \textwidth]{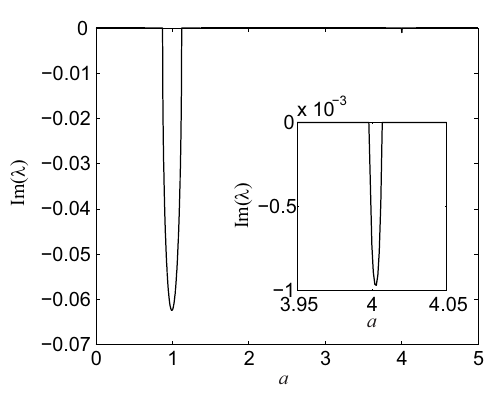}
      \caption{Imaginary part of the  exponent $\mathrm{Im}\,(\lambda)$ as a function
of $a$ for $b = 0.25$. Notice the main lobe centered around $a = 1$.
The second lobe, shown in the inset, is much smaller than
the first one that it can hardly be seen on the same scale. The figure is taken from Ref.~\cite{PhysRevA.76.063609}.}
      \label{fig:pra2007}
\end{figure}

In addition to the mentioned full numerical solution of GP Eq.~(\ref{eq:GPE}), there are other approximative ways to tackle its solutions. When considering a quasi-one- dimensional setup, one can assume a Gaussian BEC wave function along the radial direction. With this approximation an effective nonpolynomial nonlinear Schr\"odinger equation can be derived \cite{nlse1, nlse2}, which is easier to address numerically \cite{PhysRevA.76.063609, PhysRevA.86.023620}. Another possibility for understanding the onset of a pattern in a driven BEC for weak driving amplitudes, is to consider linear (in)stability analysis of Eq.~(\ref{eq:GPE}) following the original Ref.~\cite{PhysRevLett.89.210406}.
In Refs.~\cite{PhysRevLett.89.210406, PhysRevA.81.033626, PhysRevA.86.023620} a harmonic modulation of the interaction from Eq.~(\ref{eq:modulated_interaction}) has been considered.
In this part we analyse a homogeneous setup without a trap and express a possible modulated solution with wave vector ${\bf k}=(k_x, k_y, k_z)$ and time-dependent amplitudes $u(t)$ and $v(t)$ as
\begin{eqnarray}
 \psi(x,y,z,t)&& = \psi_{\mathrm{H}}(t)\times\nonumber\\ &&\hspace{-8mm}\left[1 + \left(u(t) + i v(t)\right) \cos \left(k_x x+ k_y y + k_z z\right)\right],
 \label{eq:ansatz}
\end{eqnarray}
where
\begin{equation} \psi_{\mathrm{H}}(t) = \exp\left(-i \mu t-i (\varepsilon/\omega) \sin(\omega t)\right)
\label{eq:psiH}
 \end{equation}
 is the spatially-uniform time-dependent solution and $\mu$ is the chemical potential.
By linearizing the GP equation (\ref{eq:GPE}) with respect to amplitudes $u$ and $v$ we obtain the Mathieu equation
\begin{equation}
\frac{d^{2}u}{d\tau^{2}}+\left(a({\bf k}, \omega)+b(k,\omega, \varepsilon)\cos(2\tau)\right)u=0,
\label{eq:Mathieu}
\end{equation}
where $\tau={\omega}t/2$, the coefficient $a({\bf k}, \omega)$ is set by the dispersion relation of elementary excitations $\epsilon({\bf k})$ of the system, $a({\bf k}, \omega) = 4\epsilon^2({\bf k})/\omega^2$, and the coefficient $b(k,\omega, \varepsilon)$ is given by $b(k = |{\bf k}|,\omega, \varepsilon) = 4 g \varepsilon k^2/\omega^2$.
Due to the periodicity of $\cos(2\tau)$, according to the Floquet theorem \cite{Floquet}, a generic solution of the Mathieu equation (\ref{eq:Mathieu})  takes the form
\begin{equation}
u(t)=e^{i\lambda t}f(t),
\end{equation}
where $f(t)$ has the same periodicity as $\cos(2 \tau)$,
and $\lambda$ is a complex exponent.
The region with unstable modes in the $\omega - \varepsilon$ plane is defined by the condition
\begin{equation}
 \mathrm{Im}(\lambda) < 0.
 \label{eq:imlambda}
\end{equation}
It is well known that the Mathieu equation exhibits
an intricate stability chart comprising tongues of both stable and
unstable solutions \cite{book_of_McLachlan, Mathematica}.
In the limit of weak driving amplitude $\varepsilon\rightarrow0$, the tip of the unstable tongue $a({\bf k}, \omega) = n^2, n = 1, 2, \ldots$ gives the parametric resonance condition
\begin{equation}
 \epsilon({\bf k}) = \frac{n}{2} \omega,
 \label{eq:parres}
\end{equation}
 where $n$ is a positive integer. In particular, at the first resonance $n = 1$, the condition states that the two Bogoliubov modes - elementary excitations of a BEC \cite{book_of_Pethick_Smith} - with momenta ${\bf k}$ and ${\bf -k}$  are simultaneously excited due to the conservation of momentum.
For the lobe set around $a({\bf k}, \omega)=1$, the negative part of the imaginary exponent is given by
\begin{equation}
 \mathrm{Im}(\lambda)\approx-\frac{\sqrt{b^2(k, \omega, \varepsilon)-4(a({\bf k}, \omega)-1)^2}}{4},
 \label{eq:imlambda1}
\end{equation}
where we assume that $b(k, \omega, \varepsilon)$ is small. From this, we find that the mode with the fastest growth corresponds to
\begin{equation}
a({\bf k}, \omega)\approx 1.
\label{eq:a1}
 \end{equation}
Typically the first resonance is the strongest, as shown explicitly in Fig.~\ref{fig:pra2007}.

In the region of instability defined by Eq.~(\ref{eq:imlambda}), it holds true that
\begin{equation}
  \mathrm{Re}(\lambda)=\frac{\omega}{2},
\end{equation}
which is another key feature of the parametric resonance:
on top of the exponential growth of the occupancy of the unstable mode, an oscillation with half of the driving frequency sets in.

From the resonance condition in Eq.~(\ref{eq:parres}) and from the growth rate in Eq.~(\ref{eq:imlambda1}), we infer that features of the emergent pattern depend on the system properties,  such as dimensionality, geometry of a trap, boundary conditions, and the dispersion relation $\epsilon({\bf k})$, as well as on the driving parameters - frequency and amplitude.
In an ideal case, precisely at the resonant condition even a very weak driving will lead to the onset of Faraday waves.
In this regime, the described onset of Faraday waves can be exploited as a spectroscopic tool. In particular, the approach was used experimentally to probe the dispersion relation of surface modes of a pancake-shaped BEC \cite{Phys.Rev.Lett.127.113001} and to measure the dispersion relation of a coherently coupled two-component BEC \cite{PhysRevLett.128.210401}.
However, patterns are expected for a very broad range of driving frequencies  above a certain threshold value of the driving amplitude.
In addition, within the full description given by Eq.~(\ref{eq:GPE}), nonlinearities introduce coupling to many off-resonant modes and can shift the dips of resonant tongues to finite driving amplitudes.

From Eq.~(\ref{eq:parres}) the wave number $k$ of the instability can be determined and its spatial periodicity $\lambda = 2\pi/k$. As we discuss in the next section, this approach is established as a proper way to describe Faraday waves in an elongated BEC with short-range interactions \cite{PhysRevA.76.063609}.
In an elongated trap, the linear analysis successfully provides many details of the patterns. In two dimensions the situation is more complex, as the linear stability analysis only predicts properly the initial BEC dynamics.

\section{Faraday patterns in an elongated BEC}
\label{sec:elongatedBEC}
\begin{figure}[tbhp!]
           \includegraphics[width=0.45\textwidth]{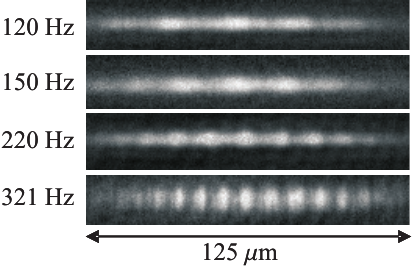}
      \caption{In-trap absorption images of Faraday waves in a BEC.
Frequency labels for each image represent the driving frequency
at which the transverse trap confinement is modulated. The figure is taken from Ref.~\cite{PhysRevLett.98.095301}.}
      \label{fig:prl2007}
\end{figure}
The first theoretical study of Faraday-pattern formation in a BEC was performed for a two-dimensional setup \cite{PhysRevLett.89.210406}. In that study harmonic modulation of the interaction strength was proposed as a method of pattern excitation. However, in the first experiment an elongated, cigar-shaped BEC was considered \cite{PhysRevLett.98.095301}. The trap parameters in the experiment were $\omega_r = 2\pi \times 160.5 \, \mathrm{Hz}, \omega_z = 2\pi\times 7 \,\mathrm{Hz}$. Faraday waves were induced by the modulation of the radial trap strength and it was suggested that the induced periodic change in density results in the effective modulation of the interaction.  The frequency of the radial breathing mode is close to $\omega_B\approx2\, \omega_r$. A range of driving frequencies $\omega\in2\pi\left[100, 400\right] \mathrm{Hz}$ was considered. For a resonant driving $\omega\approx \omega_B$ a weak driving amplitude $\varepsilon\sim 0.036$ was sufficient for visible onset of a pattern. Away from the resonance, a stronger driving amplitude $\varepsilon\sim 0.425$ was used.

The original experimental result is presented in Fig.~\ref{fig:prl2007}.
The wave vector of the induced density pattern clearly depends on the driving frequency. It was shown that except for $\omega\approx \omega_r$ the periodicity of the pattern decreases with the driving frequency $\omega$.
The onset of the Faraday wave was related to the transfer of the absorbed energy into the high-energy longitudinal modes with frequencies $\omega \sim\sqrt{l (l + 3)}, l = 1, 2, \ldots$.

The first experimental study attracted lot of interest in the topic and several follow up
theoretical papers focused on the analysis of the reported experimental data. Here we present some of the results of a
theoretical analysis from Ref.~{\cite{PhysRevA.76.063609}} where the actual experimental parameters were taken into account.
In accordance with the experimental setup of Ref.~\cite{PhysRevLett.98.095301},
the transverse frequency was modulated, as defined in Eq.~(\ref{eq:modulated_trap}).
We then present analytical results for the wave vector of the emergent pattern originally presented in Ref.~\cite{PhysRevE.84.056202}.

\begin{figure*}[!htb]
\includegraphics[width=0.8\textwidth]{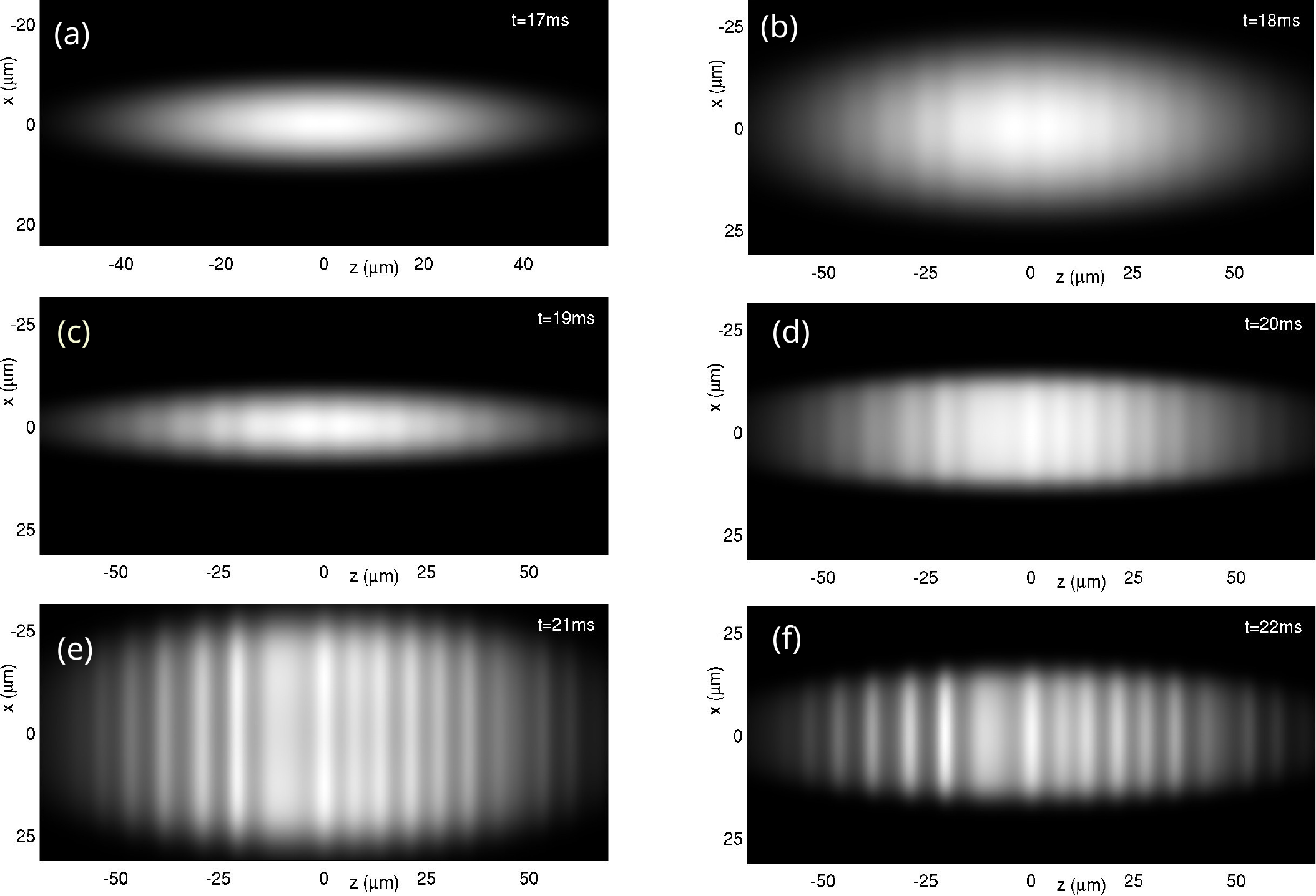}
\caption{Faraday pattern from the simulations of GP Eq.~(\ref{eq:GPE}).
This case corresponds to the
experiment of Ref.~\cite{PhysRevLett.98.095301}, namely, a cloud
of $N=5 \times 10^5$ $^{87}$Rb atoms, trapped by
$\omega_{r} = 2\pi \times 160.5$~Hz,$\omega_z = 2\pi \times 7$~Hz
with a 20\% modulation of the radial confinement at
a driving frequency $\omega/(2\pi)=321$~Hz.
The panels show snapshots of the $y$-integrated  density profile
(i.e., the observable in the experiments) at (a) $t=17\mathrm{ms}$, (b) $t=18\mathrm{ms}$, (c) $t=19\mathrm{ms}$, (d) $t=20\mathrm{ms}$, (e) $t=21\mathrm{ms}$, and (f) $t = 22\mathrm{ms}$.
The figure is taken from Ref.~\cite{PhysRevA.76.063609}.
}
\label{fig:fig3d1}
\end{figure*}

In the study of Ref.~{\cite{PhysRevA.76.063609}} several methods were used. In particular, an effective one-dimensional equation for the dynamics along the $z$-axis was derived starting from the GP Eq.~(\ref{eq:GPE}). In the derivation, it was assumed that the features of the condensate wave function in the orthogonal $x-y$ directions are captured by a time-dependent Gaussian wave function~\cite{nlse1, nlse2, PhysRevA.76.063609}. Within this approach a good agreement with the experimental data was obtained for almost all driving frequencies except for $\omega = \omega_r$.
 A full agreement with experimental data was reached when using the numerical simulation of the full GP Eq.~(\ref{eq:GPE}).  Exemplary numerical results are presented in Fig.~\ref{fig:fig3d1}.  For the particular results, the driving frequency was set at the frequency of the radial breathing mode $\omega  =2\omega_r$. The presented density profiles clearly show the onset of Faraday waves in numerical simulations.
The pattern was found to be visible at approximately $t = 18 \mathrm{ms}$ for the driving frequency $\omega = 2\omega_r$. The length scale of the observed pattern was found to be in good agreement with experimental data.

Another possibility to derive analytical results for the properties of emergent Faraday waves is to use a time-dependent variational analysis of the solutions of GP Eq.~(\ref{eq:GPE}) \cite{PhysRevA.56.1424, PhysRevLett.83.1715}, as done in Ref.~\cite{PhysRevE.84.056202}.
The three-dimensional variational wave function was decomposed
into a radial and a longitudinal part as
\begin{eqnarray}
\psi(r, z, t)&=&\mathcal{N}(k, t)\exp\left[-r^{2}/(2w^{2}(t))+i r^2 \alpha(t)\right]\nonumber\\&\times& (1 + (u(t) + i v(t))\cos(k z)),
\label{eq:decomposition_var}
\end{eqnarray}
with a temporally variable width characterized by
$w(t)$ and a conjugated phase $\alpha(t)$.
By using this ansatz in the time-dependent variational approach of Ref.~\cite{PhysRevA.56.1424}, the following equations of motion were derived:
\begin{eqnarray}
 \ddot{w}(t) &=& \left[\frac{1}{w^3(t)}+\frac{\rho g}{2\pi w^3(t)}-\omega_r^2(1+\varepsilon \sin\omega t)^2 w(t)\right],\label{eq:Ermakov}\\
 \ddot{u}(t) &=&-\frac{k^2}{2}\left[\frac{k^2}{2}+\frac{\rho g}{\pi w^2(t)}\right] u(t),
 \label{eq:Ermakov2}
\end{eqnarray}
where $\rho$ is the linear density of the condensate.
The equation (\ref{eq:Ermakov}) had been originally introduced in 1880 by Ermakov \cite{Ermakov1, Ermakov2}. It is well known that it exhibits a series of parametric resonances at $\omega = 2\omega_r, \omega_r, \omega_r/2,\ldots, 2\omega_r/n$, where $n$ is a positive integer.  The equations (\ref{eq:Ermakov}) and (\ref{eq:Ermakov2}) explicitly show the coupling of radial motion with the onset of the Faraday wave along the axial direction.

For a nonresonant driving, the second order derivative in Eq.~(\ref{eq:Ermakov}) can be disregarded and the condensate width can be approximated by
\begin{equation}
 w(t)\approx\frac{1}{\sqrt{\omega_r(1+\varepsilon \sin \omega t)}}\left(1+\frac{\rho g}{2 \pi}\right)^{1/4}.
\end{equation}
With this result Eq.~(\ref{eq:Ermakov2}) takes the form of the Mathieu Eq.~(\ref{eq:Mathieu}). The coefficients $a(k,\omega)$ and $b(k,\omega, \varepsilon)$ from Eq.~(\ref{eq:Mathieu}) can be expressed as \cite{PhysRevE.84.056202}
\begin{eqnarray}
&&a(k,\omega)=\frac{k^{4}}{\omega^{2}} + \frac{2 \rho g k^2 \omega_r}{\omega^2 \sqrt{\pi^2+\rho g \pi/2}},\nonumber\\
&&b(k,\omega, \varepsilon)=\frac{2\varepsilon \rho g k^2 \omega_r}{\omega^2\sqrt{\pi^2+\rho g \pi/2}}.\nonumber
\end{eqnarray}
The dispersion relation of Faraday waves from Eq.~(\ref{eq:a1}) was found to be \cite{PhysRevE.84.056202}
\begin{eqnarray}
k_F&=&\sqrt{\frac{\sqrt{2\rho^2 g^2 \omega_r^2+\pi \omega^2(2\pi+\rho g)}-\sqrt{2}\rho g \omega_r}{\sqrt{2\pi^2+\rho g \pi}}},
\label{eq:kF}
\end{eqnarray}
where $\rho$ is the linear density of the condensate. For the resonant driving $\omega = \omega_r$, it was shown that the resonance $a(k, \omega) = 2^2$ plays a role, and the dispersion relation of the resonant wave was found to be
\begin{equation}
 k_R = \sqrt{\frac{\sqrt{\rho^2 g^2 \omega_r^2+2\pi \omega^2(2\pi+\rho g)}-\rho g \omega_r}{\sqrt{\pi^2+\rho g \pi/2}}}.
 \label{eq:kR}
\end{equation}

\begin{figure}[t!]
\includegraphics[width=0.42\textwidth]{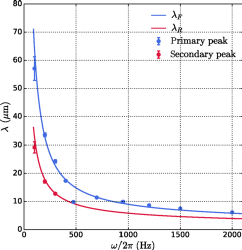}
\caption{Spatial period vs. driving frequency $\omega$. The experimental data are
indicated by filled squares, while the solid lines are the theory of
Ref.~\cite{PhysRevE.84.056202}.
The blue data points are the primary
peak of the fast-Fourier transform of the line-density profiles, while the red data points correspond to a
secondary peak, when one exists.
The error bars here correspond to the standard error of the mean determined from ten different experimental runs for each point.
The solid blue line is the calculated spatial period $\lambda_F = 2\pi/k_F$ of the
Faraday mode from Eq.~(\ref{eq:kF}), while the red is that of the resonant mode $\lambda_R = 2\pi/k_R$ from Eq.~(\ref{eq:kR}).
The figure is taken from Ref.~\cite{PhysRevX.9.011052}.}
\label{fig:figprx1}
\end{figure}
\begin{figure}[t]
\centering
\includegraphics[width=0.44\textwidth]{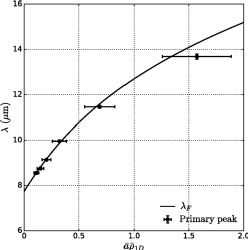}
\caption{ Interaction dependence of $\lambda_F = 2\pi/k$. The experimental data are
indicated by filled squares, while the solid line is the theory of
Ref.~\cite{PhysRevE.84.056202}. The figure is taken from Ref.~\cite{PhysRevX.9.011052}.}
\label{fig:figprx2}
\end{figure}

These analytical results were compared to the experimental measurements reported in Ref.~\cite{PhysRevX.9.011052} and a perfect agreement was found for a wide range of driving frequencies and interaction strengths, as shown in Figs.~\ref{fig:figprx1} and \ref{fig:figprx2}. We note that the comparison is possible although the presented derivation is given for the modulated trap while in the experiment the interaction strength was modulated, as the applied driving mechanism does not play the key role here \cite{JPhysSocJpn.92.064602}. In the experiment a gas of up to $8\times 10^5$ Li atoms was confined in a harmonic trap with frequencies $\omega_r = 2\pi \times 475 \, \mathrm{Hz}, \omega_z = 2\pi\times 7 \,\mathrm{Hz}$. A wide range of driving frequencies $\omega\in2\pi\left[100, 2000\right] \mathrm{Hz}$ was considered. In the experiment column-density
distributions were measured and integrated along the $y$ axis to obtain
line-density profiles. A fast-Fourier transform was applied to these profiles in order to determine the spectrum
of spatial frequencies exhibited by the BEC following
modulation. In the experimental data, there were two peaks found in the spatial distribution for the driving frequencies below $\omega\leq2\pi \times 500 \mathrm{Hz}$. These two peaks were denoted as primary (corresponding to $k_F$) and secondary (corresponding to $k_R$). Except at the exact resonance, it was found that the spatial period decreases with increasing driving frequency $\omega$, see Fig.~\ref{fig:figprx1}. Also, the dependence of the spatial period on the effective nonlineartity is found to be in perfect agreement with the prediction from Eqs.~(\ref{eq:kF}) and (\ref{eq:kR}).
Finally, for a slower driving $\omega\ll\omega_r$, Faraday pattern was not observed. In this regime a strong driving amplitude was applied leading to the fragmentation of the initial BEC, which then turns into a state with irregular granulated density distribution. This process is not captured by the GP equation. By using the multiconfigurational time-dependent Hartree method \cite{RMP.92.011001} it was shown that the granulated state exhibits strong quantum fluctuations and correlations.

\section{Patterns in dipolar BECs}

The  dipole-dipole interaction (DDI) between atoms or molecules is anisotropic and long-range. These features are responsible for a whole series of new phenomena that appear in ultracold dipolar gases~\cite{PhysRep.464.71}. Additionally, the study of dipolar cold atoms has attracted considerable attention following the realization that quantum fluctuations in these systems can stabilize novel states of matter \cite{FrontPhys.16.32202, Rep.Prog.Phys.86.026401, NatRPhys.5.735}. In the context of pattern formation, the dispersion relation of a dipolar BEC develops a minimum at finite momentum \cite{PhysRevLett.90.250403, PhysRevLett.122.183401}. This distinctive feature makes it particularly relevant to investigate how Faraday wave characteristics are modified in the presence of DDI. The results presented in this section are based on analytical and numerical analyses, while experimental confirmation remains an open challenge.

The dipolar Gross-Pitaevskii equation takes the form
   \begin{eqnarray}
 i\hbar\frac{\partial\psi}{\partial t}&=&\left[-\frac{\hbar^{2}}{2m}\nabla^{2}
+V(x, y, z, t)+g(t) N\left|\psi\right|^{2}\right.\nonumber\\&&\left.+ N \int d{\bf r}'\, \psi^*({\bf r}')
       U_\mathrm{dd}({\bf r - r}', t) \psi({\bf r}') \right]\psi,
\label{eq:GPEdipolar}
\end{eqnarray}
where the dipolar potential reads $U_\mathrm{dd} ({\bf r}) = 3a_{\mathrm{dd}}(1 - 3 \cos^2 \theta)/r^3$, $\theta$ is the angle between the dipoles' orientation  and vector ${\bf r}=(x, y, z)$ connecting the two dipoles, and $a_\mathrm{dd}$ is the DDI interaction strength, that depends on the dipole moment of atoms $d$ and their mass $m$ as $a_\mathrm{dd} = \mu_0 m d^2 / (12 \pi \hbar^2)$, where $\mu_0$ is the vacuum permeability. In Eq.~(\ref{eq:GPEdipolar}) we explicitly write a possible time dependence of DDI which may be implemented
with intensity oscillations of the polarizing electric field for the case of polar molecules,
or with additional transverse magnetic fields, which lead to a precession of the dipole
moment orientation, for the case of magnetic dipoles \cite{PhysRevLett.89.130401, PhysRevLett.120.230401}. We also assume that all dipoles are oriented parallel to each other.

In the context of dipolar BECs, the study of Faraday waves was first considered for the case of harmonically modulated long-range interaction in one-dimensional and two-dimensional systems~\cite{PhysRevA.81.033626, PhysRevA.86.023620}.
In the following we present results obtained for a parametric modulation of the dipole-dipole interactions
\begin{equation}
g_{d}(t) = \bar{g}_d (1+2 \varepsilon \cos(2 \omega t)),
\label{eqn:gdModulationFormula}
\end{equation}
where $\varepsilon$ characterizes the modulation strength.  For a highly elongated BEC, a strong trap along the radial direction was assumed, while no trap was taken into account along the $z$ axis. The magnetic dipole moment was assumed to be oriented by an external ﬁeld along the $y$ axis. In this quasi one-dimensional setup, the DDI are determined by the coupling constant
$\bar{g}_{d} = 3a_{\mathrm{dd}} n_{0}/(2 \pi l_r)$,
where $l_r =\sqrt{\hbar/m\omega_r}$.

By starting from the nonlocal nonlinear Schr\"odinger equation for a quasi one-dimensional BEC, derived from Eq.~(\ref{eq:GPEdipolar}), and by linearizing the equation of motion with the ansatz of Eq.~ (\ref{eq:ansatz}) where $k_x = 0, k_y =0, k_z \neq0$, the following Mathieu equation can be derived
\begin{equation}
\frac{d^{2}u}{dt^{2}} + \left[ \epsilon^{2}(k) +
  2 \omega^{2} b(k,\omega,\alpha) \cos(2 \omega t) \right] u = 0.
\label{eqn:mathieuSingle}
\end{equation}
In the last equation, the dispersion relation of elementary excitations of the dipolar BEC reads
\begin{equation}
\epsilon(k) = \sqrt{ \frac{k^2}{2} \left( \frac{k^2}{2} +
  2 g + \frac{4\pi}{3} \bar{g}_d F_0(k) \right) },
\label{eq:bogolibovSingleTube}
\end{equation}
where the DDI term yields the factor
$F_{0}(k) = 1 + \frac{3}{2} k^{2} e^{k^2/2} \mathrm{Ei}\left(-k^2/2\right)$,
and $\mathrm{Ei}(x)$ is the exponential integral function. The coefficient $b$ from the Mathieu equation takes the form
\begin{equation}
b(k,\omega,\varepsilon) = \frac{2\pi}{3 \omega^{2}} \bar{g}_d \varepsilon k^{2} F_0(k).
\label{eqn:defB}
\end{equation}
\begin{figure}[t]
\includegraphics[width=0.45\textwidth]{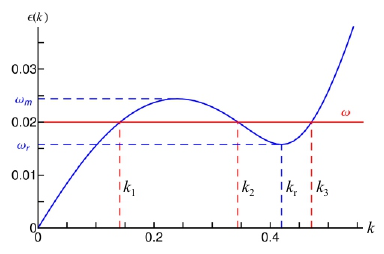}
\caption{Excitation spectrum $\epsilon(k)$ given by Eq.~(\ref{eq:bogolibovSingleTube})
of a single quasi-one-dimensional dipolar BEC, with $g=-0.1007$, and
$\bar{g}_{d}=0.0629$. Note the roton minimum at $\omega_r=\epsilon(q_r)$
and the maxon maximum at $\omega_m$. For a driving
frequency $\omega_r<\omega<\omega_m$ there are three
possible momenta $k_{1,2,3}$ obeying the resonance
condition $\epsilon(k)=\omega$. The figure is taken from Ref.~\cite{ PhysRevA.86.023620}.
}
\label{fig:dipolar1}
\end{figure}
The spectrum of
elementary excitations in the considered case is presented in Fig.~\ref{fig:dipolar1}. In addition to the sound mode at low momenta, it exhibits a maximum at $\omega_m$ and a minimum at $\omega_R$, and it is known as roton-maxon spectrum \cite{PhysRevLett.90.250403, PhysRevLett.122.183401}. The roton minimum signals a tendency toward the formation of crystalline order. For the range of driving frequencies $\omega_R<\omega<\omega_m$, there are three values of momenta $k_1, k_2, k_3$ satisfying the parametric resonance condition (\ref{eq:parres}). By comparing the instability growth rates it turns out that the most unstable mode occurs at the largest momenta $k_3$. This result is different from the closely related result for a quasi-two-dimensional BEC \cite{PhysRevA.81.033626} where the intermediate momentum $k_2$ has the largest negative imaginary part of the instability exponent.

\begin{figure*}[t]
\includegraphics[width=0.9\textwidth]{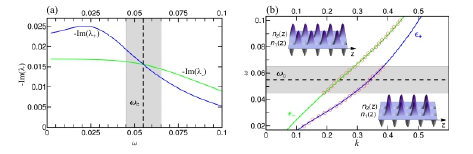}
\caption{
(a) Negative imaginary part $-\mathrm{Im} (\lambda_\pm)$ of
the Floquet exponent, corresponding to the the first
parametric resonance for the symmetric and anti-symmetric
excitation branches $\epsilon_{\pm}(k)=\omega$, as a
function of the driving frequency $\omega$. Note that
at a critical frequency $\omega_{c}=0.055$, both exponents
are equal, $\mathrm{Im} (\lambda_+) = \mathrm{Im} (\lambda_-)$, indicating a transition
between the symmetric and the anti-symmetric Faraday pattern.
Other parameters correspond to $g=-0.0435$, $\bar{g}_d =0.0437$,
$D=6l_\perp$, and $\varepsilon=0.02$.
(b) Analysis of the pattern selection
as a function of the driving frequency $\omega$, in the
neighborhood of the critical frequency $\omega_{c}$ .
The
solid lines represent the excitation branches. For each
$\omega$  a circle indicates a momentum value where
the numerical Fourier transform $\hat{n}_{j}(k_{z})$
of the Faraday pattern shows a clear maximum. For $\omega$
well below~(above) $\omega_{c}$  we observe a single peak
at $k_+$~($k_-$) indicating that a symmetric~(anti-symmetric)
Faraday pattern emerges~(see insets). In the vicinity of
$\omega_{c}$ (shaded region), both modes are equally unstable
and the two corresponding peaks occur
simultaneously in the Fourier transform (see text). The figures are taken from Ref.~\cite{ PhysRevA.86.023620}.
}
\label{fig:dipolar5}
\end{figure*}

Implications of the long-range DDI are even stronger for the onset of Faraday waves in two coupled quasi-one-dimensional BECs. In Ref.~\cite{PhysRevA.86.023620} it was assumed that two BECs are confined in two parallel one-dimensional traps at a distance $D$ from each other. In this case there is no tunnelling of atoms between the two BECs which couple to each other exclusively through long-range interaction.
These non-local
interactions lead to a collective character of the elementary excitations
that are shared among the two quasi-1D condensates~\cite{PhysRevA.85.033618}.
Consequently, the excitation spectrum unfolds into two branches
\begin{equation}
\epsilon_{\pm}(k) = \sqrt{ \frac{k^2}{2} \left ( \frac{k^2}{2} + 2 g + \frac{4\pi}{3} \bar{g}_d
\left ( F_0(k) \pm F_1(k) \right ) \right) },
\label{eqn:bogolibovDoubleTubes}
\end{equation}
which correspond, respectively, to symmetric and anti-symmetric states
with respect to the transposition of traps $j=1\leftrightarrow j=2$. In the last equation the DDI introduces factors $
F_{p}  \left( k_z \right) = \! \int\limits_{0}^{\infty} \!dk_r \frac{k_r e^{-\frac{1}{2}k_r^{2}}}{k_r^{2}+k_{z}^{2}}
\,  \left[ \left(k_r^{2}-2 k_{z}^{2}\right) J_{0} \left(k_r D\, p \right) -3 k_r^{2} J_{2}\left( k_r D \,p  \right) \right],
$
where $J_{n}(x)$ are the Bessel functions of the first kind, the index $p$ takes values $0$ or $1$ and $D$ denotes the distance between the two coupled BECs.

In this case, by considering the symmetric and anti-symmetric combinations $u_\pm =u_1\pm u_2$ of the $u_{1}$ and $u_2$ amplitudes of the two BECs, two independent Mathieu equations were derived:
\begin{equation}
\frac{d^{2}u_{\pm}}{dt^{2}} + \left[ \epsilon^{2}_{\pm}(k)
  + 2 \omega^{2} b_{\pm}(k,\omega,\varepsilon)
\cos(2 \omega t) \right] u = 0,
\label{eqn:mathieuMagic}
\end{equation}
with
\begin{equation}
b_{\pm}(k,\omega,\varepsilon) = \frac{2\pi}{3 \omega^{2}} \bar{g}_d \varepsilon k^{2} \left ( F_0(k)\pm F_1(k) \right ).
\end{equation}
As in the case of a single BEC, the
first parametric resonances $\epsilon_\pm (k_\pm)=\omega$ are characterized
by the Floquet exponents $\mathrm{Im}(\lambda_\pm) \simeq b_{\pm}(k,\omega,\varepsilon)/2
\propto k^{2} ( F_0(k)\pm F_1(k) )$, and the emerging Faraday pattern is
determined, for each driving frequency separately, by the mode with the
largest $-\mathrm{Im}(\lambda)$. Remarkably, the involved momentum dependence of
$F_0(k)\pm F_1(k)$ leads to an intricate relation between the Floquet
exponents and the driving frequency $\omega$, as presented in Fig.~\ref{fig:dipolar5}. A critical frequency $\omega_c$ was found such that the maxima of the two emergent Faraday waves in the two BECs align with each other for driving frequencies below $\omega_c$. On the contrary,  for the driving frequencies $\omega > \omega_c$  the maxima of the first of the Faraday waves align with the minima of the second wave. The two cases are illustrated in the insets of Fig.~\ref{fig:dipolar5}(b).

Finally, we review results from a recent study of dipolar BECs \cite{Symmetry.11.1090} with harmonically modulated radial trap strength given by equation (\ref{eq:modulated_trap}). In this study, an elongated trap was considered, where the condensate is confined into a cigar-shaped harmonic trap, with the equilibrium frequencies $\omega_{r} = 2 \pi \times 160.5$~Hz and $\omega_{z} = 2 \pi \times 7$~Hz. These are typical values taken from Ref.~\cite{PhysRevLett.98.095301}. The dipole moments of the atoms were assumed to be oriented along the $x$ direction, i.e., orthogonal to the weak-confinement axis $z$ (which we refer to as the longitudinal axis), since this maximizes the stability of the system. The condensate is considered to have $N=10^4$ atoms.

For the full simulations of Eq.~(\ref{eq:GPEdipolar}) the programs described in Refs.~\cite{ComputPhysCommun.180.1888,ComputPhysCommun.183.2021,ComputPhysCommun.195.117,ComputPhysCommun.200.406,ComputPhysCommun.200.411,ComputPhysCommun.204.209,ComputPhysCommun.220.503,ComputPhysCommun.209.190,ComputPhysCommun.240.74} were used. The parameters of these simulations match the physical parameters of BECs of chromium $^{52}$Cr \cite{Nature.448.672}, erbium $^{168}$Er \cite{PhysRevLett.108.210401}, and dysprosium $^{164}$Dy \cite{PhysRevLett.107.190401}, which, respectively, have the dipole moments $d=6\mu_\mathrm{B}$, $d=7\mu_\mathrm{B}$, and $d=10\mu_\mathrm{B}$,
where $\mu_\mathrm{B}$ is the Bohr magneton. The~corresponding background $s$-wave scattering lengths are $a_s=105 a_0$, $a_s=100 a_0$, and $a_s=100 a_0$, where $a_0$ is the Bohr radius.
The driving frequency was typically set at the value $\omega = 2 \pi  \times  200$~Hz.
To characterize the Faraday waves, their Fourier transformed spectra in the time-frequency and     spatial-frequency domains were analyzed.

\begin{figure*}[t]
    \centering
 \includegraphics[width=0.8\textwidth]{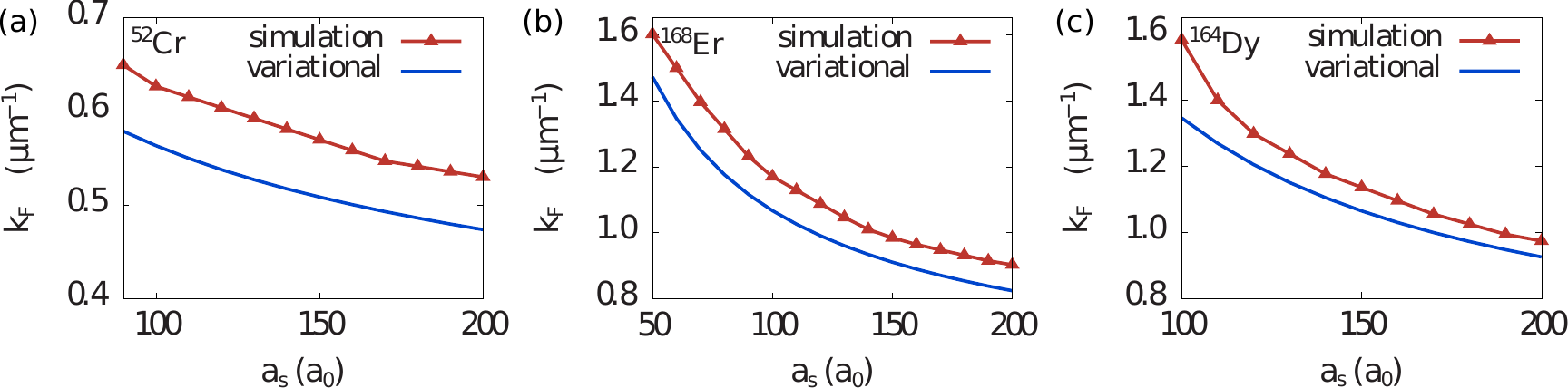}
    \caption{Wave vector of the Faraday waves $k_F$ as a function of the contact interaction strength for a BEC of  (a) $^{52}$Cr, (b)
     $^{168}$Er, and (c) $^{164}$Dy, for a fixed DDI strength. Red upper triangles refer to the values obtained using Fourier analysis of the results obtained by the numerical solution of Eq.~(\ref{eq:GPEdipolar}), and blue lines are the variational results according to Equation (\ref{eq:k_F}). The figure is taken from Ref.~\cite{Symmetry.11.1090}.}
    \label{fig:faraday_k_F_acc}
\end{figure*}
\begin{figure*}[ht]
    \centering
 \includegraphics[width=0.8\textwidth]{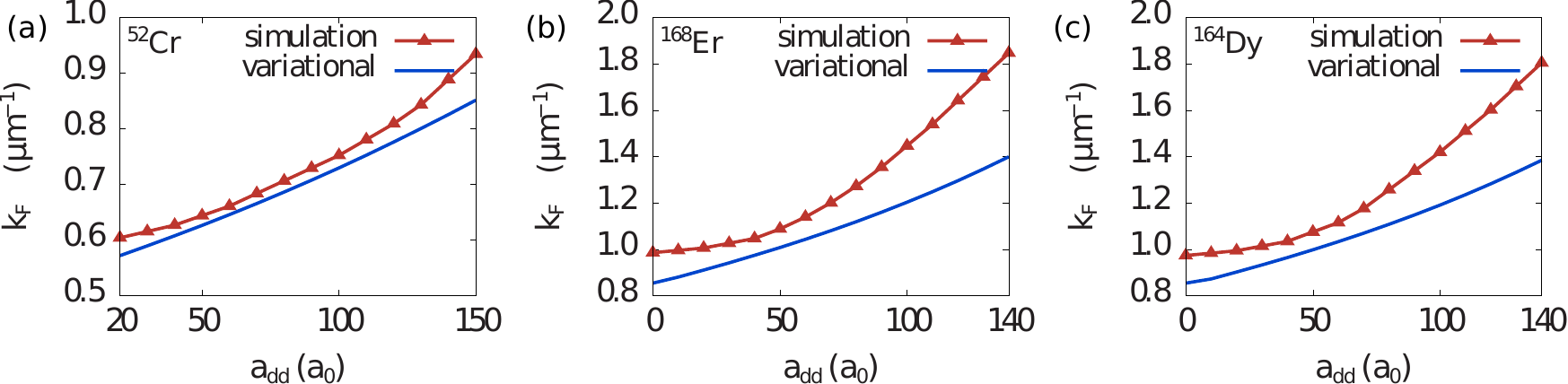}
    \caption{Wave vector of the Faraday waves $k_F$ as a function of the DDI strength for a BEC of (a) $^{52}$Cr, (b)
     $^{168}$Er, and (c) $^{164}$Dy, for a fixed contact interaction strength. Red upper triangles refer to the values obtained using Fourier analysis of the results obtained by the numerical solution of Eq.~(\ref{eq:GPEdipolar}), and blue lines are the variational results according to Equation (\ref{eq:k_F}). The figure is taken from Ref.~\cite{Symmetry.11.1090}.}
    \label{fig:faraday_k_F_add}
\end{figure*}
In order to gain further insight into the properties of the observed patterns, a mean-field variational approach for the dynamics of a driven dipolar BEC at zero temperature was developed by using the ansatz of Eq.~(\ref{eq:decomposition_var}) with $k_x = 0, k_y = 0, k_z \neq 0$.
This yields the expressions for the coefficients $a(k, \omega)$ and $b(k, \omega)$ of the resulting Mathieu equation (\ref{eq:Mathieu})
\begin{equation}
    a(k, \omega) =  \frac{k^4}{\omega^2} + \frac{ \Lambda k^2}{\omega^2} \, , \qquad b(k, \omega, \varepsilon) = \frac{ \Lambda k^2 \varepsilon}{\omega^2} \, ,
\end{equation}
where $\Lambda$ is given by
\begin{equation}
    \Lambda = \frac{4 \sqrt{\frac{2}{\pi}} \, N \left[ a_s
        - \frac{a_\mathrm{dd}}{2} \, f_s \left( \frac{R_ \rho}{R_ z} \right) \right]}
        {R_ z \left\{ 1 + \sqrt{\frac{2}{\pi}} \frac{N}{R_ z} \left[
            a_s + \frac{a_\mathrm{dd}}{2} f_s \left( \frac{R_ \rho}{R_ z} \right)
            - a_\mathrm{dd} f_s^{\prime}\left( \frac{R_ \rho}{R_ z} \right)
        \right] \right\}^{1/2} } \, ,
\end{equation}
with $f_s(x) = f(x, x)$, $f(x, x)$ being the standard dipolar anisotropy function~\cite{PhysRevA.74.013621},
 and $R_ \rho$ and $R_ z$ are ground-state condensate sizes in radial and longitudinal directions.

As previously discussed, the instability condition for the Faraday waves reads $a(k, \omega) = 1$, Eq.~(\ref{eq:a1}), which can be used to calculate the wave vector of Faraday waves shortly after their emergence,
\begin{equation}
    k_{F} = \sqrt{- \frac{\Lambda}{2} + \sqrt{\frac{ \Lambda^2}{4} + \omega^2}} \, .
    \label{eq:k_F}
\end{equation}
The analytically obtained results for the spatial period of Faraday waves are compared to results of the extensive numerical simulations, which solve the full three-dimensional GP Eq.~(\ref{eq:GPEdipolar}) for a dipolar BEC and a very good agreement was found, as shown in Figs.~\ref{fig:faraday_k_F_acc} and \ref{fig:faraday_k_F_add}.  Figure~\ref{fig:faraday_k_F_add} presents the corresponding dependence of $k_F$ on $a_\mathrm{dd}$ for a fixed value of the contact interaction. In contrast to the contact interaction dependence, where $k_F$  is a decreasing function of $a_s$, here we see that $k_F$ increases as the DDI strength is increased.
Overall, variational results capture the observed features of $k_F$ as a function of interaction strength, and the quantitative agreement of the variational approximation with numerically exact results is reasonable.

\section{Surface patterns in a pancake BEC}

While the onset of Faraday waves in an elongated BEC has been investigated from various sides since early days, progress in understanding Faraday waves in a two-dimensional setup was more gradual. One of the milestones in the study of pattern formation in pancake shaped BECs was the understanding of dynamical instabilities related to the surface waves  \cite{Phys.Rev.Lett.127.113001, Phys.Rev.A.102.033320}.

\begin{figure*}[ht]
\centering
\includegraphics[width=0.8\textwidth]{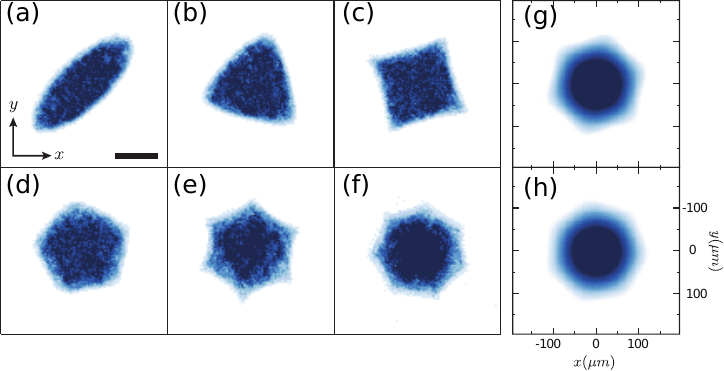}
\caption{ Experimental observation of star-shaped condensates. (a-f) Representative intrap images (single-shots) of the ensuing regular polygons with $D_l$ symmetry triggered by the periodic modulation of the s-wave scattering length.
The modulation frequencies are 84~Hz ($D_2$), 104~Hz ($D_3$), 119~Hz ($D_4$), 132 Hz ($D_5$), 147~Hz ($D_6$), and 161~Hz ($D_7$), respectively. The background scattering length is set to $a_{bg} = 138(6) a_B$, and the mean modulation amplitude is $\bar{a}_m=19a_B$, where $a_B$ is the Bohr radius.
(g) Hexagon and (h) heptagon shaped patterns obtained by solving the  GP Eq.~(\ref{eq:GPE}). The figure is taken from Ref.~\cite{Phys.Rev.Lett.127.113001}.}
\label{fig:prl2021_Fig1}
\end{figure*}
In the experiment reported in Ref.~\cite{Phys.Rev.Lett.127.113001} a harmonically trapped system with trapping frequencies  $\omega_r = 2\pi\times29.4(2),\omega_z=2\pi\times725(5)$~Hz was considered. A Feshbach resonance was used for the modulation of the s-wave scattering length of $^{7}$Li, see also Section 6.1. Depending on the value of the driving frequency, within relatively narrow frequency regions, an oscillating star-shaped BEC cloud with $l$-fold symmetry was observed, as presented in Fig.~\ref{fig:prl2021_Fig1}. Under the harmonic modulation, a radially symmetric BEC turned into an oscillating regular polygon with random orientation. In  Fig.~\ref{fig:prl2021_Fig1} we observe different possible shapes - an ellipse, a triangle, a square, a pentagon and so on.

\begin{figure*}
\includegraphics[width=0.8\linewidth]{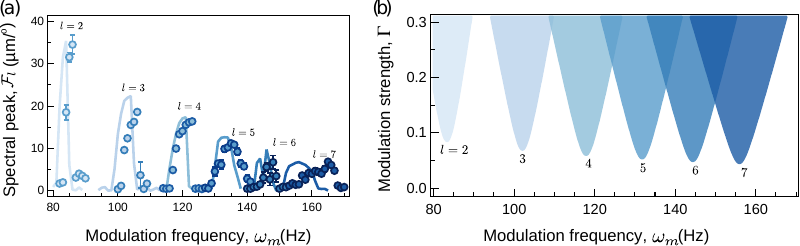}
\caption{ Hydrodynamic instability of the surface excitations. (a) Spectral peak of various $l$-fold star patterns, created after 1~s of modulation, as a function of the modulation frequency.
The filled circles designate the experimental results and solid lines refer to the GP equation predictions; notice the very good agreement between the two.
(b) Floquet stability tongues for different modulation strengths $\bar{a}_m/a_{bg}$ and frequencies $\omega_m/(2\pi)$ characterizing distinct $l$-fold patterns. The figure on the left can be interpreted as the intersection of the instability tongues from the right figure at the driving amplitude $\Gamma=a_m/a_{bg} = 0.14$. The figure is taken from Ref.~\cite{Phys.Rev.Lett.127.113001}.}
\label{fig:prl2021_Fig2}
\end{figure*}

A related Mathieu equation was derived
\begin{equation}\label{eq:surfMathieu}
\ddot{\zeta}_{l}(t) + \omega^2_l\big[1+ \frac{\bar{a}_m}{a_{bg}}\cos(\omega_m t)\big]\zeta_{l}(t) = 0,
\end{equation}
where $\omega_l = \sqrt{l} \omega_r$ are the natural frequencies of the surface modes of the harmonically trapped BEC. To trigger this pattern formation a density disturbance of the form $\delta n = \zeta_l(t)r^{l}e^{il\phi}$ was assumed. The predictions of the derived effective Mathieu equation were confirmed by comparison with the simulations of the full GP Eq.~(\ref{eq:GPE}) as well as with experimental data, see Figs.~\ref{fig:prl2021_Fig1} and \ref{fig:prl2021_Fig2}. To achieve full agreement a phenomenological dissipative term of the form $\gamma \dot{\zeta}_{l}(t)$ was added to Eq.~(\ref{eq:surfMathieu}). The role of this term was to introduce a finite threshold value of the driving amplitude above which the instability sets in.
In the numerical simulations, a weak random amplitude was added on top of the BEC ground state to mimic a low thermal fraction of atoms.
The nonlinear effects are observed in the asymmetry of the peaks found in the Fourier spectra as a function of the driving frequencies. The asymmetries get smaller with weaker driving amplitudes and the onset of surface waves at weak driving amplitudes was used to probe the frequencies of surface modes with great accuracy of $2\%$. Furthermore, the measurements demonstrated the exponential growth of the occupancies of the most unstable mode, which is another important feature of a parametrically driven system.

Along these lines, it was proposed in Ref.~\cite{Phys.Rev.A.102.033320} that the properties of surface modes at the boundary between the two immiscible BECs can be probed through harmonic modulation of the scattering length of one of the species leading to the onset of surface waves.

\section{Stabilization of a pattern in a two-dimensional BEC}

The linear stability analysis presented in the previous sections \ref{sec:methods} and \ref{sec:elongatedBEC} captures properties of an emergent Faraday pattern for an elongated BEC quite well in a broad range of physical parameters, as confirmed experimentally \cite{PhysRevLett.98.095301, PhysRevA.76.063609, PhysRevX.9.011052}. However, the situation is more complex in two dimensions. In this case the resonant condition (\ref{eq:parres}) puts a constraint on the modulus of the wave vector, but this property does not fix all the features of the emergent pattern \cite{PhysRevLett.89.210406}. In the initial dynamical stage, both experimental results and full numerical simulations  show the onset of stripes in random directions, in agreement with the condition (\ref{eq:parres}). In this regime  the two Bogoliubov modes with ${\bf k}$ and ${\bf -k}$  are simultaneously excited due to the momentum conservation law. For very strong driving amplitudes these early excitations are proven to be precursors for the formation of atomic jets that show up in the later evolution stage \cite{PhysRevLett.121.243001, Nature551.356}.


A response of a disc-shaped cesium BEC under three different modulation schemes of the scattering length was investigated in Ref.~\cite{NatPhys.16.652}. Initially, the BEC is weakly interacting and its dispersion relation is well approximated by the noninteracting dispersion. Afterwards a strong driving of the interaction is applied.
In the first scheme a single frequency $\omega$ was used which resulted in the onset of stripe patterns with $k_F =  \sqrt{m\omega/\hbar}$,  where $m$ is the atomic mass. Within the second scheme, the interaction strength was modulated at frequency $\omega$ in the seeding stage and then a second frequency $\omega/2$ was superposed in the pattern forming stage
This scheme resulted in a hexagonal lattice pattern. Finally, in the third scheme at early stages the frequency $\omega/2$ was used and then switched to frequency $\omega$. This scheme gave rise to the formation of a square lattice.

In order to describe the emergent density wave, it is necessary to go beyond the linear stability analysis and to
take into account both the growth in the mode occupancies as well as a growth suppression due to coupling of different modes emanating from intrinsic nonlinearities of the model. This approach was used for the derivation of the respective quantum nonlinear amplitude equations \cite{NatPhys.16.652}.  The momentum and energy conservation of the bosonic stimulated scattering processes from the BEC were taken into account to explain the onset of different spatial structures \cite{NatPhys.16.652}.

\begin{figure*}[tb]
\centering
\includegraphics[width=0.8\textwidth]{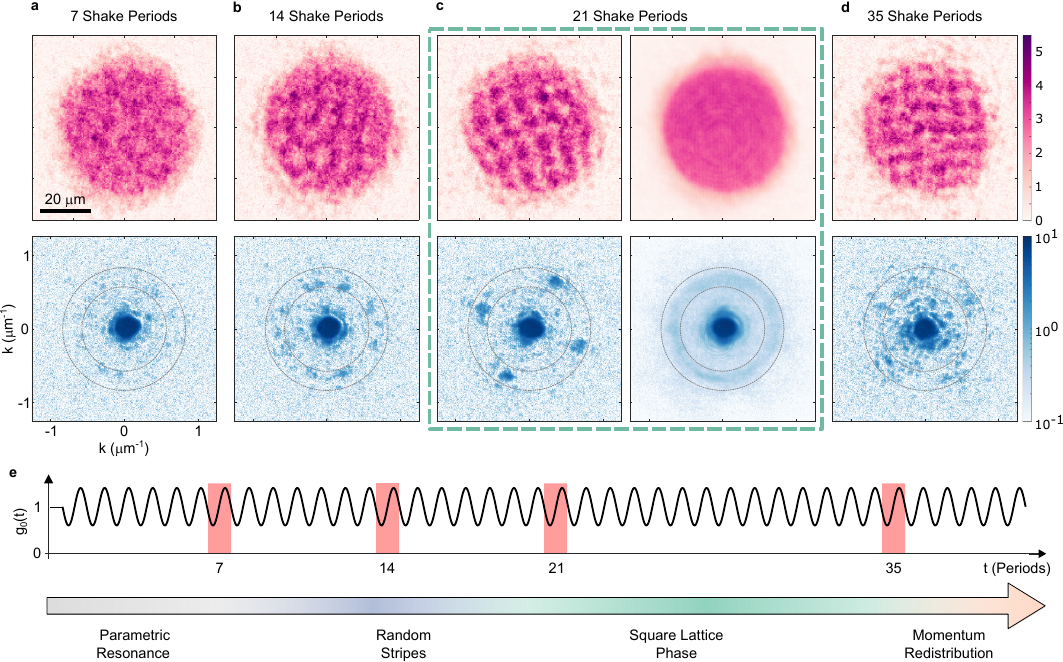}
\caption{(a) Real space (top row) and momentum space (bottom row) distributions after 7 shake periods with driving amplitude  $\varepsilon = 0.4$ and driving frequency $\omega = 2\pi \times 400\mathrm{Hz}$. The dashed lines in momentum space show the region of resonant momenta, centered at $k_c = 2\pi \times 0.11$\,$\mu\mathrm{m}^{-1}$. (b) Distributions after 14 driving periods, where density modulations have become more apparent in real space and appear in momentum space as back-to-back correlated, randomly oriented peaks.  (c) Structures in the square lattice phase. The left column shows single realizations, whereas the right column shows averaged real and momentum space distributions. The smooth mean distributions indicate that structures are formed spontaneously in random directions. (d) Late times show that square lattices at the characteristic length scale are still apparent, but other momenta also become occupied. Both color codes indicate the signal in atoms per pixel. (e) The interaction is periodically modulated with a single frequency and a non-zero offset. The colored arrow indicates the various phases during time evolution. The figure is taken from Ref.~\cite{PhysRevX.15.011026}.}
\label{fig:prx2025}
\end{figure*}

More recently, for the purpose of addressing the limit of a homogeneous two-dimensional atomic cloud, a specifically tailored trapping potential was used. In the experiments recently reported in  Refs.~\cite{PhysRevA.109.L051301, PhysRevX.15.011026, NatPhys.21.1064} an atomic cloud of $^{39}$K was confined within a disc with flat potential, while a potential barrier with a linear slope was implemented at the disc boundaries. In this way, the reflection of phononic excitations from the boundaries was minimised. In this setup, it was experimentally shown that the BEC exhibits an emergent square-lattice pattern under harmonic driving of interaction. An example of the experimental result from Ref.~\cite{PhysRevX.15.011026} is presented in Fig.~\ref{fig:prx2025}. As shown before, stripe patterns emerge in the early driving stages, while at intermediate time scales a randomly oriented square lattice is stabilized.

In order to explain these features, amplitude equations for an interacting BEC \cite{PhysRevLett.89.210406} were extended and refined \cite{PhysRevA.109.L051301, PhysRevX.15.011026} using the multiple-time-scale method \cite{PhysRevLett.89.210406, 0204406}. Here we only briefly draft the main steps in the derivation for the two distinct modes with wave vectors ${\bf k}$ and ${\bf p}$.
In the multiple-time-scale method a bookkeeping parameter $\delta$ is introduced, such that $\varepsilon\sim \delta^2, (\omega/2-\epsilon({\bf k}))\sim\delta^2, (\omega/2-\epsilon({\bf p}))\sim\delta^2$.  A slow time scale is defined as $\tau = \delta^2\,t$ and all time-dependent functions become dependent on two variables $f(t)\rightarrow f(t, \tau)$, $\partial_t \rightarrow \partial_t + \delta^2 \partial_{\tau}$.
The condensate wave function is also expanded  in terms of $\delta$ on top of the homogeneous solution $\psi_H(t)$ from Eq.~(\ref{eq:psiH}). By employing the aforementioned reparametrization and performing an order-by-order expansion in terms of $\delta$, GP Eq.~(\ref{eq:GPE}) transforms in the hierarchy of differential equation where high-order terms depend on the low-order terms. Convergence of this expansion is ensured by requiring that the secular terms vanish in the equations for higher-order terms in the perturbative expansion. This requirement imposes constraints on the leading-order solution.

To the leading order in $\delta$ the time-dependent wave function is given by
\begin{equation}
 \psi(t, \mathbf{r}) = \psi_{\mathrm{H}}(t) \left[1 +  \phi_{\mathbf{k}}(t)\cos(\mathbf{k} \cdot\mathbf{r}) + \phi_{\mathbf{r}}(t)\cos(\mathbf{p} \cdot\mathbf{r})\right],
 \label{eq:expansion}
\end{equation}
where the time-dependent factors can be expressed in terms of Bogoliubov amplitudes
\begin{eqnarray}
 \phi_{\mathbf{k}} (t)&=& \left(1-\frac{E_0(\mathbf{k})+2\mu}{\epsilon(\mathbf{k})}\right) R(t, \mathbf{k}) e^{i\omega t/2}\nonumber\\ &+& \left(1+\frac{E_0(\mathbf{k})+2\mu}{\epsilon(\mathbf{k})}\right) R^*(t, \mathbf{k}) e^{-i \omega t/2}.
\end{eqnarray}
In the last equation $\epsilon(\mathbf{k}) = \sqrt{ E_0({\mathbf k}) \left(E_0({\mathbf k}) + 2\mu\right)}$ is the Bogoliubov dispersion relation, $E_0({\mathbf k})$ is the noninteracting dispersion relation and $\mu = g n$ is the chemical potential. Starting from the time-dependent GP Eq.~(\ref{eq:GPE}) and using the described multiple-scale method, it was shown that the complex amplitudes $R(t, \mathbf{k})$ obey a complex  Ginzburg-Landau type of equation
\begin{eqnarray}
 &&\hspace{-4mm}i\frac{d}{d t}  R(t, \mathbf{k}) = \Delta R(t, \mathbf{k}) + i\alpha R^*(t, \mathbf{k})\nonumber\\&&\hspace{-4mm}-i \Gamma R (t, \mathbf{k})
 +\lambda \left[|R(t, \mathbf{k})|^2 R(t, \mathbf{k})\right.\nonumber\\
 &&\hspace{-4mm}\left.+ c_1(\theta)|R(t, \mathbf{p})|^2 R(t, \mathbf{k})+c_2(\theta)R(t, \mathbf{p})^2 R^*(t, \mathbf{k})\right],
 \label{eq:amp-eq-with-D}
\end{eqnarray}
where $\Delta = \omega/2-\epsilon(\mathbf{k})$ is the detuning, $\alpha = \mu \varepsilon E_0(\mathbf{k})/(2 \epsilon(\mathbf{k}))$ is the driving amplitude for the Bogoliubov mode, and $\lambda = \mu \left(5 E_0(\mathbf{k}) + 3\mu\right)/\epsilon(\mathbf{k})$ is a nonlinearity. The coefficients $c_1$ and $c_2$ depend on the angle $\theta$ between the vectors $\mathbf{k}$ and $\mathbf{p}$.
Their values were determined from the requirement that the secular terms vanish in the equations for higher-order terms in the perturbative expansion.
Finally, a phenomenological dissipative term proportional to $\Gamma$
was introduced to effectively capture the dynamics beyond the two modes of interest. This term leads to irreversible evolution of the reduced system, even though the underlying GP equation is reversible. However, it was shown that the magnitude of $\Gamma$ is irrelevant as long as the parameters allow for the occurrence of the Faraday instability at $\alpha>\Gamma$.

\begin{figure}[tb]
\centering
\includegraphics[width=0.45\textwidth]{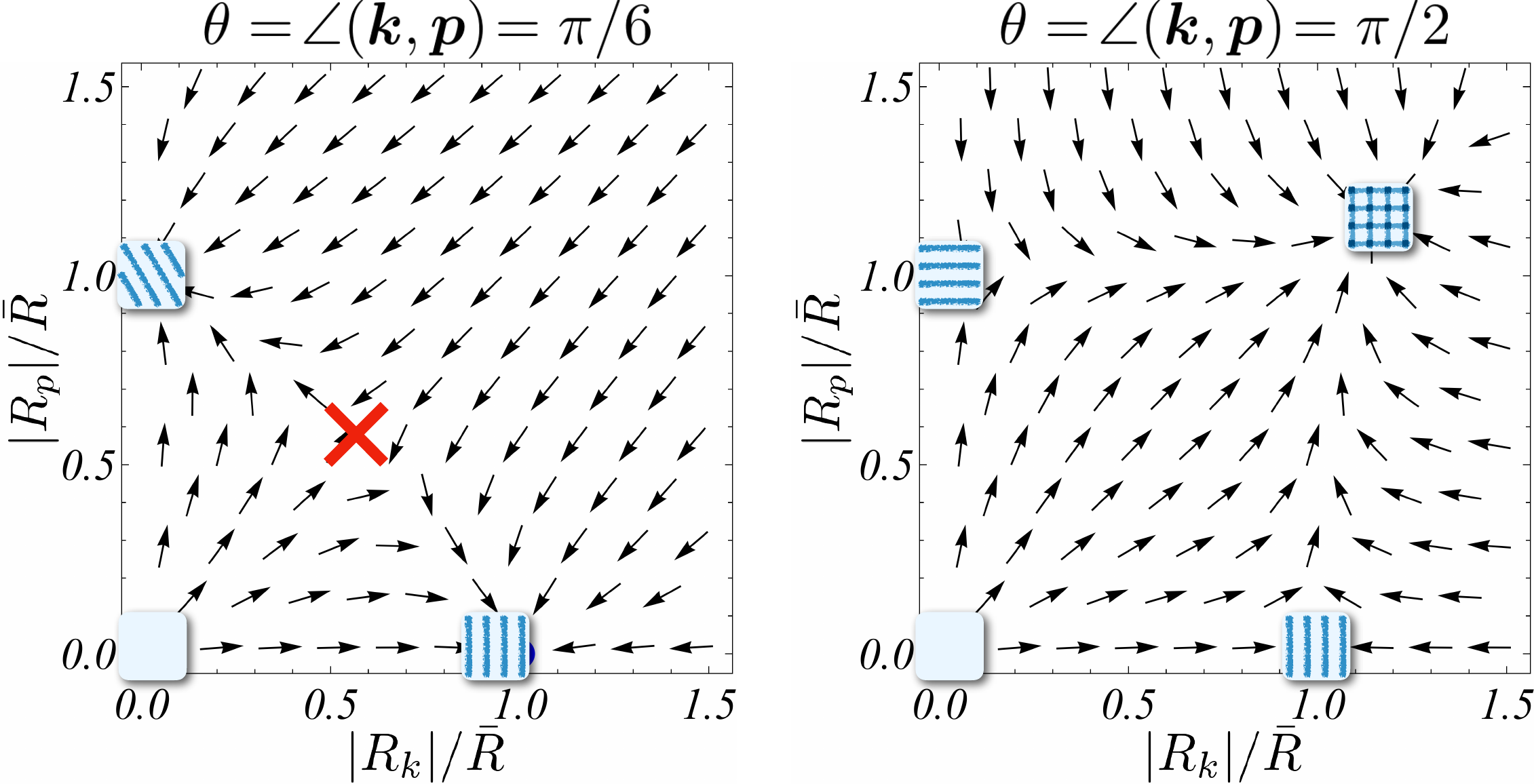}
\caption{
Global stability of the patterns formed by two standing waves in a planar BEC in directions $\mathbf{k}$ and $\mathbf{p}$.
Schematic figures at the fixed points represent corresponding stationary solutions, i.e., the grid-pattern, stripe-pattern, and uniform solutions.
The angle $\theta$ between the two excited modes is $\pi/6$ (left) and $\pi/2$ (right). In the latter case, the square grid pattern emerges as a stable fixed point.
The parameters are $\omega/\mu=2$ and $\varepsilon=0.6$ with low dissipation $\Gamma=0.1\alpha$, where $\omega,\mu$ and $\alpha$ are the driving frequency, the chemical potential, and the drive amplitude, respectively.
The figure is taken from Ref.~\cite{PhysRevA.109.L051301}.}
\label{fig:pra1}
\end{figure}

By setting $d R(t, \mathbf{k})/d t=0$ for the resonant driving $\Delta = 0$ in Eq.~\eqref{eq:amp-eq-with-D}, such that $k = p$, four possible fixed-point values of $R_{k}$ and $R_{p}$ were found:
\begin{subequations}
\begin{align}
(\textrm{U})\quad& R_{k}=R_{p}=0, \\
(\textrm{S}_k)\quad&
R_{k}=\bar{R}e^{i\bar{\eta}},\quad
R_{p}=0, \label{eq:fixed-pi-2a}\\
(\textrm{S}_p)\quad& R_{k}=0,\quad R_{p}=\bar{R}e^{i\bar{\eta}}, \\
(\textrm{G})\quad& R_{k}=R_{p}
=\bar{R}e^{i\bar{\eta}}/\sqrt{1+c_{1}(\theta)+c_{2}(\theta)},
\label{eq:grid-amp}
\end{align}
\end{subequations}
with $\bar{R}^{2}=\sqrt{\alpha^{2}-\Gamma^{2}}/\lambda$ and $\exp(i\bar{\eta})=(\sqrt{\alpha-\Gamma}+i\sqrt{\alpha+\Gamma})/\sqrt{2\alpha}$.
The fixed points correspond to the following density patterns: (U) a uniform pattern, (S$_k$) and (S$_p$)  stripe patterns for each of the direction, and (G) a grid pattern. By a thorough analysis of local and global stability of the fixed points, as presented in Fig.~\ref{fig:pra1}, it was shown that the grid structure is stable for $\theta = \pi/2$, in agreement with experimental observations.

Properties of these emergent lattice structures have been further probed by introducing different type of defects in a system in a controlled way. In this way, it was shown that the collective modes of the driven superfluid with the emergent density-wave state exhibits some of the key features of a supersolid state \cite{NatPhys.21.1064}.

\section{Conclusions}
In this chapter, we have presented several selected theoretical and experimental milestones relevant for the understanding of the process of pattern formation in driven condensates. Much of the theoretical understanding of the phenomenon is based on the derivation of an effective Mathieu equation from the underlying Gross-Pitaevskii equation and on the features of the parametric resonance originating in the emergent effective equation. For this reason, we first reviewed some basic properties of the Mathieu equation. We then discussed the case of Faraday-wave formation in an elongated BEC.
Several experiments had addressed the onset of Faraday waves through harmonic modulation of trap strength and through harmonic modulation of local interaction for a wide range of driving frequencies and driving amplitudes in an elongated BEC \cite{PhysRevLett.98.095301, PhysRevX.9.011052, PhysRevLett.121.185301, PhysRevLett.128.210401}.  We presented an efficient analytical derivation based on the time-dependent variational calculation that captures many of the observed features, such as spatial periodicity of the emergent pattern and instability growth rates \cite{PhysRevE.84.056202}. For low driving frequencies and strong driving amplitudes, a process of granulation occurs, which is beyond the validity of the Gross-Pitaevskii equation.

Long-range dipolar interactions enrich the phenomenon of pattern formation. We have reviewed three theoretical proposals which are still awaiting experimental confirmation. The essence of the predicted features comes from the way the complex dispersion relation of a dipolar BEC affects the onset of parametric resonance \cite{PhysRevA.81.033626, PhysRevA.86.023620}. The effect is particularly pronounced in the case of two BECs coupled by long-range interactions. Finally, we presented an analytical result for the periodicity of a Faraday wave of an elongated harmonically-trapped BEC with dipolar interactions induced by the periodic modulation of the radial harmonic trap strength \cite{Symmetry.11.1090}.

Following the progress toward realization of Faraday waves in two-dimensional settings, we reviewed the onset of surface waves in a pancake shaped BEC induced through harmonically driven local interaction. Depending on the driving frequency, an oscillating polygon-shaped BEC had been found in remarkable agreement with the theory based on an effective Mathieu equation \cite{Phys.Rev.Lett.127.113001}.
Finally, we discussed stabilization of a lattice in a two-dimensional setup. We first reviewed the result obtained by bichromatic driving of local interactions \cite{NatPhys.16.652}. Next, we analyzed the onset of robust density wave states forming a square lattice under harmonic driving of local interaction. We drafted the derivation of amplitude equations that capture long-time BEC dynamics in this case and predict a  square-lattice solution as a stable fixed point \cite{PhysRevA.109.L051301,PhysRevX.15.011026} .

Overall, the results included in this review evidence that in the past two decades through mutual experimental and theoretical progress much has been learned about pattern formation in a driven BEC. In essence, the ideas from the first theoretical  proposal are now enriched, refined and realized in various experimental setups. This progress and the high degree of control over pattern formation in driven BECs allows and requires further research on the emergence of exotic states under driving, in particular in three dimensions. The onset of Faraday waves in systems with long-range interactions and spin-orbit coupling is awaiting its first experimental verifications. The reported contribution of quantum fluctuations to the dynamics of a strongly driven BEC \cite{PhysRevX.9.011052} and the observation of entanglement in parametrically excited collective modes of a BEC \cite{PhysRevLett.135.240603} open the door for the realization of strongly correlated states. Strong driving with a uniform force has been shown to lead to the onset of turbulence \cite{Nat.620.521}, although the role of the pattern formation in this process remains to be explored. In  a broader context, pattern formation in dissipative-driven systems \cite{RMP.97.025004}, where it is anticipated to facilitate the characterization of emergent nonequilibrium steady states.

{\bf Acknowledgement} This preprint will appear as a chapter in the Springer book entitled \textit{Short and Long Range Quantum Atomic Platforms — Theoretical and Experimental Developments} (provisional title), edited by P. G. Kevrekidis, C. L. Hung, and S. I. Mistakidis. The authors acknowledge funding provided by the
Institute of Physics Belgrade through a grant from the
Ministry of Science, Technological Development, and Innovation of the Republic of Serbia. For this work M.C.R. and A.N.-\.{Z} were supported by the Romanian Ministry of Education and Research under Romanian National Core Program LAPLAS VII— contract no. 30N/2023.


\end{document}